\documentstyle{elsart}
\input psfig
\def\lsim{\mathrel{\rlap{\lower4pt\hbox{\hskip1pt$\sim$}}
    \raise1pt\hbox{$<$}}}         
\def\gsim{\mathrel{\rlap{\lower4pt\hbox{\hskip1pt$\sim$}}
    \raise1pt\hbox{$>$}}}         
\begin{document}

\begin{frontmatter}

\title{Do $hep$ neutrinos affect the solar neutrino energy spectrum ?}
\author{{John N. Bahcall}\thanksref{jnbemail}} 
\address{School of Natural Sciences, 
Institute for Advanced Study, Princeton, NJ 08540}
\author{Plamen I. Krastev\thanksref{pikemail}}
\address{Department of Physics, University of Wisconsin, Madison, WI 53706}
\thanks[jnbemail]{E-mail address: jnb@ias.edu}
\thanks[pikemail]{E-mail address: krastev@nucth.physics.wisc.edu}

\begin{abstract}
If the low energy cross section for ${\rm ^3He} ~+~ p 
~\rightarrow ~{\rm ^4He} ~+~e^+ ~+~\nu_e $, the `$hep$' reaction, is 
$\gsim 20$ times larger than the best (but uncertain) 
theoretical estimates, then
this reaction could significantly influence the  electron energy 
spectrum produced by  solar neutrino interactions and measured 
in the SuperKamiokande, SNO, and 
ICARUS experiments.
We compare predicted energy spectra for different assumed $hep$
fluxes and different neutrino oscillation scenarios 
with the observed SuperKamiokande spectrum. The spectra with enhanced
hep contributions provide better fits to the SuperKamiokande data.
\end{abstract}

\end{frontmatter}

\section{Introduction}

One of the primary science goals of the 
SuperKamiokande~\cite{superkamiokande504,superkamiokande300}, SNO~\cite{sno}, and
ICARUS~\cite{icarus} solar neutrino experiments is to determine  the 
shape of the solar
neutrino spectrum between $\sim 5$ MeV and $14$ MeV. In this 
energy range, neutrinos from the $\beta$-decay  of 
 $^8$B  are  expected, according to solar model 
calculations carried out using 
the best available 
nuclear physics data, to dominate
the solar neutrino spectrum~\cite{bahcall89}.  
The shape of the neutrino energy spectrum from a single
$\beta$-decaying source is independent of all solar physics to an
accuracy of $1$ part in $10^5$~\cite{bahcall91}. 
Hence, a measurement of the shape is a
direct test of whether something happens to the solar neutrinos after
they are created, \hbox{\it i. e.,} of the 
minimal standard electroweak model.

The SuperKamiokande collaboration has
provided~\cite{superkamiokande504} preliminary 
data for the energy distribution  of recoil electrons created by solar
neutrinos scattering off electrons in their detector. The data are
presented in $15$ bins between $6.5$ MeV and $14$ MeV and one
higher-energy bin, $14$ to $20$ MeV, for a total of $16$ bins. The
three highest energy bins show a relatively  large number of events,
more than would have been expected from the most popular neutrino
oscillation parameters discussed prior to the first detailed 
report of the energy spectrum by the SuperKamiokande
collaboration~\cite{superkamiokande504}.

Could this excess of high energy events be caused by  $hep$
neutrinos, which have an endpoint well beyond the $\sim 14$ MeV 
endpoint of the
$^8$B energy spectrum?

The $hep$ reaction, first discussed in connection with solar neutrinos
by Kuzmin~\cite{kuzmin},

\begin{equation}
{\rm ^3He} ~+~ p ~\rightarrow ~{\rm ^4He} ~+~e^+ ~+~\nu_e ,
\label{hepreaction}
\end{equation}
produces neutrinos with an endpoint energy of $18.8$ MeV, the
highest energy expected for solar neutrinos. It was pointed out about 
a decade ago~\cite{bu88} 
that solar neutrino detectors that  measure individual
recoil electron energies, like
SuperKamiokande~\cite{superkamiokande504}, 
SNO~\cite{sno}, and ICARUS~\cite{icarus},  might be able to 
detect the  $hep$ neutrinos.
The total rate is expected to be low, but 
the  background  is small in this energy range.

The goal of this paper is to demonstrate the sensitivity of inferences
regarding the distortion of the solar neutrino energy spectrum to
assumptions regarding the low energy cross section factor,
$S_0({\rm hep})$, for the hep
reaction ( 
Eq.~[\ref{hepreaction}]),  and to emphasize the importance of experimental
and theoretical studies of  the possible contribution  of $hep$ neutrinos.

\section{Solar model calculations}
\label{s:solarmodel}

For a given solar model, the flux of $hep$ neutrinos can be calculated
accurately 
once the low-energy cross section factor
for reaction~(\ref{hepreaction}) is specified.
The rate of the $hep$ reaction is so slow that it does not affect 
solar model  calculations.
Using the uncertainties given in Ref.~\cite{bp98} for 
the solar age, chemical
composition, luminosity, radiative
opacity, diffusion rate, and for all    
nuclear quantities except $S_0({\rm hep})$, 
 we calculate a  total uncertainty in the $hep$ flux of
only $3$\% if the $S-$factor is known exactly.
The best-estimate  $hep$ 
flux is very small~\cite{bp98}: 

\begin{equation}
\phi({\rm hep}) = 2.1(1.0 \pm 0.03) 
[ { { S_0({\rm hep})} \over {2.3  \times 10^{-20} {\rm keV~ b} } } ] 
\times 10^3~{\rm cm^{-2}s^{-1}}.
\label{eq:hepflux}
\end{equation} 
The bracketed-factor in Eq.~(\ref{eq:hepflux}) is equal to unity for
the currently-recommended value for $S_0({\rm hep})$ (see discussion
in the following section). 

The best-estimate $^8$B neutrino flux is more than $2000$ times larger
than the flux of $hep$ neutrinos given by   Eq.~(\ref{eq:hepflux})
if the bracketed factor is set equal to unity.
This is the reason why all previous 
discussions of the measurement of the
energy spectrum in the SuperKamiokande, SNO, and ICARUS experiments
have concentrated on the recoil electrons produced by $^8$B neutrinos.
Even with the high event rate of SuperKamiokande($\sim 6800$ solar
neutrino events observed in $504$ days), only a  few $hep$ interactions
are expected for the standard estimate of $S_0({\rm hep})$.

\section{Calculated $hep$ Production Cross Sections}

Table~\ref{table:cross} lists all the published values of the low-energy cross
section factor, $S_0({\rm hep})$, with which we are familiar. 
Since reaction~(\ref{hepreaction}) occurs via the  weak interaction, 
the cross section
for $hep$ neutrinos is too small (see Eq.~\ref{eq:hepvalue} below) to  be
measured in the laboratory at low enough energies to be relevant to
solar fusion 
and must therefore be calculated theoretically.
We have also given in Table~\ref{table:cross}, 
in the column next to each cross section estimate,
 one or more  characteristic
features of the  physics that was used to estimate the cross section.
A review of the history of calculations of $S_0({\rm hep})$ provides
insight into the difficulty of obtaining an accurate value. 

The first estimate of the cross section factor by
Salpeter~\cite{salpeter} considered only the overlap of an incoming
continuum wave function with that of a bound nucleon in $^4$He,
obtaining a  large value for $S_0({\rm hep})$, $\sim 300$ times the
current best-estimate.  However, Werntz 
and Brennan~\cite{werntz67}
pointed out that if one approximates the final $^4$He state by
$(1s)^4$ and the initial state by $(1s)^3(s_c)$ (where $s_c$ is a
continuum initial state), and antisymmetrizes in space, spin, and
isospin, then the matrix element 
of the usual allowed $\beta$-decay operator vanishes between the
initial and final states.  They obtained a cross section factor more
than two orders of magnitude smaller than the single-particle Salpeter
estimate. 

Werntz and
Brennan~\cite{werntz67,werntz73} derived and used in an exploratory
way a suggested
proportionality between the $\beta$-decay matrix element, $M_\beta$
(which cannot be measured), 
for the reaction $^3$He(p,$e^+~\nu_e )^4$He, and the neutron-capture 
matrix element (which can be measured), 
$M_\gamma$, for the reaction $^3$He(n,$\gamma )^4$He. 
Their derivation, which was intended only to give a crude estimate of
the cross section factor, 
neglected initial state interactions and  the small $D-$state
contributions of the $^3$He and $^4$He ground states
and also assumed the dominance of meson
exchange currents.

Tegn\'{e}r and Bargholtz~\cite{tegner} stressed the importance of the
$D$-state components of the $^3$He and $^4$He wave functions and
argued  that the matrix elements for nucleon capture on $^3$He 
are dominated by one-body operators
rather than the two particle meson exchange terms.  They derived a
new  proportionality relation between $M_\beta$ and $M_\gamma$,
which they used to estimate a rather large range of possible values
for $S_0({\rm hep})$. 
Wolfs et al.~\cite{wolfs} and Wervelman et al.~\cite{wervelman} 
measured accurately the thermal neutron capture rate for
$^3$He(n,$\gamma )^4$He and  
used
the proportionality relation of Tegn\'{e}r and Bargholtz to estimate
values of $S_0({\rm hep})$.

Carlson~\cite{carlson} revealed another layer of complexity by
performing a detailed calculation with sophisticated wave functions,
showing the presence in their model of strong destructive interference
between the mesonic exchange currents and the one-body matrix elements
connecting the small components of the wave functions. In the most
comprehensive calculation to date, Schiavilla et al.~\cite{schiavilla}
included a more consistent treatment of the $\Delta$-isobar current
and investigated the sensitivity of $S_0({\rm hep})$ to the assumed
details of the nuclear physics. They found a range of values 

\begin{equation}
S_0({\rm hep}) = (2.3 \pm 0.9) \times 10^{-20} {\rm keV~ b},
\label{eq:hepvalue}
\end{equation}
which corresponds to a fusion cross section of $\sim 10^{-50} \ {\rm
cm^2}$ at solar thermal energies.
The central value of this range, $S_{0,~{\rm cent.}}({\rm hep})$,
was adopted by Adelberger et
al.~\cite{adelberger} and Bahcall and Pinsonneault~\cite{bp98} as the
best available estimate.
A value of  $S_0({\rm hep})$ in the range $20-30$  times $S_{0,~{\rm
cent.}}({\rm hep})$
 would be consistent (see discussion in the
following section) with all the available evidence from solar neutrino
experiments. 

Is it possible to show from first-principle physics that 
$S_0({\rm hep})$ cannot exceed, e. g.,  $10$ times 
$ S_{0,~{\rm cent.}}({\rm hep})$?  We have been 
unable to find any such argument. 
Therefore, for the last decade we have not quoted  a total uncertainty in
the calculated standard model predictions for the $hep$ neutrino
fluxes, although well-defined total uncertainties are given for all of the
other fluxes~\cite{bu88,bp98}.

The reason
it is difficult to place a firm upper limit 
is, as emphasized by Carlson et al.~\cite{carlson} and Schiavilla et
al.~\cite{schiavilla}, that the calculated value of $S_0({\rm hep})$ is
sensitive to the model used to describe both the ground sate
and the continuum wave functions and to the detailed form of the two-body
electroweak interactions. The matrix element $M_\beta$ contains
separate contributions from both the traditional single particle 
Gamow-Teller operator and the axial exchange-current
operator. In the most detailed calculations~\cite{carlson,schiavilla},
there is a delicate cancellation between   comparable
contributions from the one-body and the two-body operators.  For
example, if one artificially changes the sign of the principal
exchange current contribution relative to the sign of the one-body
axial current in the calculation described in Table~III of
Ref.~\cite{carlson}, the size of the calculated  $S_0({\rm hep})$ is
increased by a factor of $32$.

For non-experts, 
it is instructive to compare
the calculations of the $pp$ and $hep$ reactions.  The $pp$ fusion
reaction~\cite{kamionkowski} occurs via the 
allowed Gamow-Teller $\beta$-decay matrix element whereas the $hep$
transition is forbidden. 
For the $pp$ reaction,
the difficult-to-evaluate mesonic exchange
corrections and matrix elements connecting small components of the
wave functions are only small corrections ($\sim$ a few percent)
to the total cross section.
For the $hep$ reaction, the exchange corrections and matrix elements
involving small components of the wave function are the whole story.
For the $pp$ reaction, the
effective range approximation allows one to use measured data to
calculate to good accuracy the low energy   fusion cross
section. The somewhat analogous scaling laws relating  $hep$ fusion to
the measured cross section for $^3$He(n,$\gamma )^4$He reaction are
not valid because low energy nucleon capture by $^3$He occurs via
competing and cancelling small effects and because of different
initial state interactions. Hence, the estimated uncertainty in
the low energy $pp$ fusion cross section 
is small($\sim 2$\%~\cite{adelberger}), whereas the
uncertainty in the $hep$ cross section is much larger and is difficult to
quantify.

\begin{table}[t]
\baselineskip=16pt
\centering
\begin{minipage}{3.5in}
\caption[]{Calculated values of $S_0$(hep). 
The table lists all the published values with which we are familiar 
of the low energy cross
section factor for the hep reaction shown in 
Eq.~\ref{hepreaction}.\protect\label{table:cross}}
\begin{tabular}{llcc}
\noalign{\smallskip}
\hline
\noalign{\smallskip}
$S_0$(hep)&\multicolumn{1}{c}{Physics}&Year&Reference\\
$\left(10^{-20}\ {\rm kev~b}\right)$&\multicolumn{1}{c}{}&{}&\\
\noalign{\smallskip}
\hline
\noalign{\smallskip}
630&single particle&1952&\cite{salpeter}\\
3.7&forbidden; $M_\beta \propto M_\gamma$&1967&\cite{werntz67}\\
8.1&better wave function&1973&\cite{werntz73}\\
4-25&D-states + meson exchange &1983&\cite{tegner}\\
$15.3 \pm 4.7$&measured $^3He(n,\gamma)^4He$&1989&\cite{wolfs}\\
57&measured $^3He(n,\gamma)^4He$&1991&\cite{wervelman}\\
&shell model&&\\
1.3&destructive interference,&1991&\cite{carlson}\\
&detailed wavefunctions&&\\
1.4-3.1&$\Delta$-isobar current&1992&\cite{schiavilla}\\
\noalign{\smallskip}
\hline
\noalign{\smallskip}
\end{tabular}
\end{minipage}
\end{table}

\section{Global fits to solar neutrino data}
\label{s:fitstosuperk}

We have
investigated the predicted effects 
on solar neutrino experiments of an arbitrary
size $hep$ flux, which we will parameterize by multiplying 
$ S_{0,~{\rm cent.}}({\rm hep})$ by a constant $\alpha$ that is much
greater than unity (cf. Eq.~\ref{eq:hepvalue} ),

\begin{equation}
\alpha ~\equiv~  { { S_0({\rm hep})} \over {(2.3  \times 10^{-20} {\rm keV~
b} )} } ~ . 
\label{eq:alphadef}
\end{equation}

We have carried out
global fits to all of the solar neutrino data, the measured total event rates
in the chlorine~\cite{chlorine}, GALLEX~\cite{gallex},
SAGE~\cite{sage}, and 
SuperKamiokande~\cite{superkamiokande504} experiments, the
SuperKamiokande energy spectrum~\cite{superkamiokande504}, and 
the SuperKamiokande zenith-angle dependence of the event 
rate~\cite{superkamiokande504}.  We use the methods and the data
described fully in Ref.~\cite{bks98}, hereafter BKS98 .  For MSW fits,
the degrees of freedom (d.o.f.) are: $4$ (total rates in $4$
experiments)  $ + 15 $ (normalized spectrum for $16$ bins) $+ 9 $
(normalized angular distribution for $10$ bins) $-2$ (oscillation
parameters) $-1$ (hep flux) or $25$ d.o.f.  . For vacuum
oscillations, the Day-Night asymmetry (1 d.o.f.) is a more powerful
discriminant than the zenith-angle distribution~\cite{bks98}.  Hence,
for vacuum oscillations we have $17$ d.o.f.  .
The only
substantial 
difference from BKS98 is that in the present  paper we find the best-fit 
to the measured SuperKamiokande energy
spectrum by including an arbitrary amount of $hep$ neutrinos in addition
to the conventional $^8$B flux.
The contribution of the $hep$ flux to the total event rates is
negligible for all of the best-fit solutions.

\begin{figure}[t]
\centerline{\psfig{figure=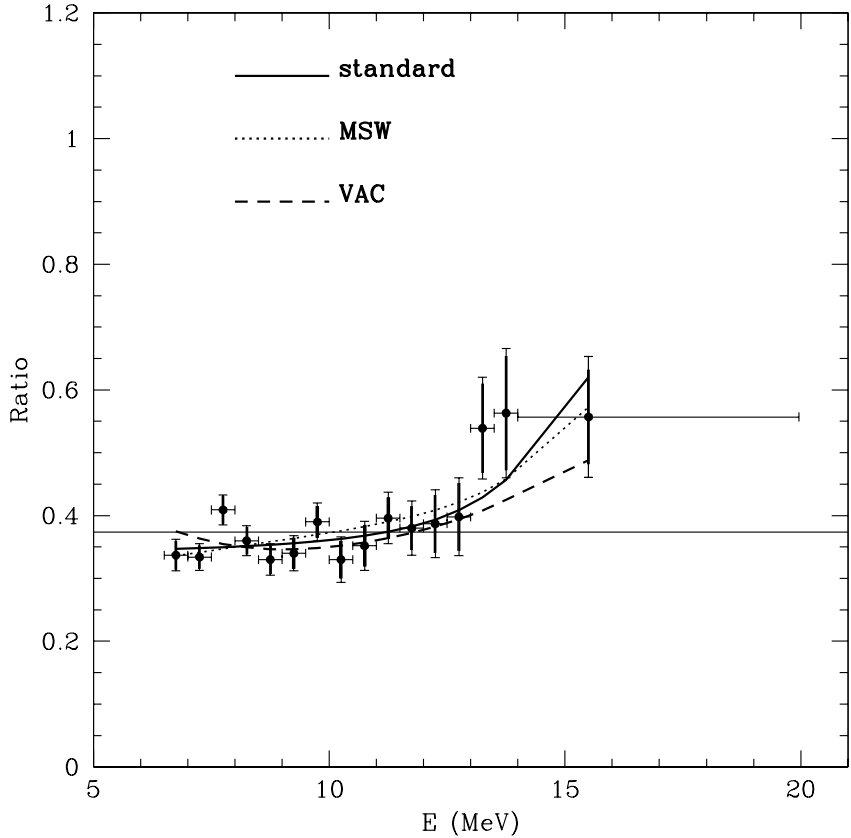,width=5in,angle=0}}
\caption{Combined $^8$B plus $hep$ energy spectrum. 
The total flux of $hep$ neutrinos was varied to obtain the best-fit
for each scenario.
The figure shows
the Ratio of the measured~\cite{superkamiokande504} to the calculated
number of events with electron recoil energy, $E$. The measured points
were reported by the SuperKamiokande collaboration at
Neutrino~98\cite{superkamiokande504}. The  calculated curves 
are global fits to all of the data, the chlorine~\cite{chlorine}, 
GALLEX~\cite{gallex},
SAGE~\cite{sage}, and SuperKamiokande~\cite{superkamiokande504} 
total event rates, the Superkamiokande~\cite{superkamiokande504}
energy spectrum, and the SuperKamiokande~\cite{superkamiokande504} 
Day-Night asymmetry. 
The calculations follow  the precepts of
BKS98~\cite{bks98} for the best-fit global solutions for a standard
 `no-oscillation'  energy spectrum, as well as MSW and
vacuum neutrino oscillation solutions.
The horizontal line at ${\rm Ratio} = 0.37$ represents the ratio of the
total event rate measured by SuperKamiokande to the predicted event
rate\cite{bp98} with no oscillations and only $^8$B neutrinos.}
\label{fig:fitstospectrum}
\end{figure}

Figure~\ref{fig:fitstospectrum} and 
Table~\ref{table:higherenergy} summarize our principal results.  We see from
Fig.~\ref{fig:fitstospectrum} that one can obtain good  fits to
the reported SuperKamiokande~\cite{superkamiokande504} energy spectrum
for all three neutrino scenarios: no oscillations, MSW, and vacuum
oscillations. 

Table~\ref{table:higherenergy} shows the best global fits to all the
data that are possible by allowing large 
enhancements of
the current best-estimate $hep$ flux.
The improvements are significant.  

\begin{table}[t]
\baselineskip=16pt
\centering
\begin{minipage}{3.5in}
\caption[]{Global fits with arbitrary $hep$ neutrino flux. The table
lists the best-fit enhancement parameters, $\alpha$, defined by
Eq.~(\ref{eq:alphadef}), for three different neutrino scenarios: no
oscillations, MSW, and vacuum oscillations. We also list the
 value of $\chi^2_{\rm min}$ for the global fit ($25$ d.o.f. for MSW
fits and $17$ d.o.f. for vacuum oscillations fits) 
and the confidence level  $P$ at which the solution is rejected, as
well as the expected number of neutrino events in the $14$-$16$ MeV
bin and the $16$-$20$ MeV bin for the $504$ day data set of
SuperKamiokande(normalized to the total number of observed 
events\cite{superkamiokande504}
$14-20$ MeV). 
The best-fit values for $\Delta m^2$ and $\sin^22\theta$ are given in
the text. For comparison, we also list the results for the best-fit
global solutions obtained in Ref.~\cite{bks98} for the standard $hep$
flux, i. e., $\alpha = 1.0$( with one less d.o.f.). 
\protect\label{table:higherenergy}}
\begin{tabular}{lccccc}
\noalign{\smallskip}
\hline
\noalign{\smallskip}
Neutrino&{$\alpha_{\rm best}$}&$\chi^2_{\rm
min}$&$P$&$14-16$ &$16-20$\\
&&&&events&events\\
\noalign{\smallskip}
\hline
\noalign{\smallskip}
No oscillations&26&25.3&0.954&62&14\\
No oscillations&1.0&36.6&0.998&70&6\\
MSW&25&30.7&0.80&64&12\\
MSW&1.0&37.2&0.93&70&6\\
Vacuum&30&23.0&0.85&66&10\\
Vacuum&1.0&28.4&0.94&69&7\\
\noalign{\smallskip}
\hline
\noalign{\smallskip}
\end{tabular}
\end{minipage}
\end{table}

The best-fit MSW solution improves from a confidence level 
 (1 - P) of $7$\% to
$20$\% (for $\alpha = 26$) even after accounting for the extra
d.o.f. \ .  
A large range of values of $\alpha$
($\lsim 30$) give good  fits. 

\begin{figure}[t]
\centerline{\psfig{figure=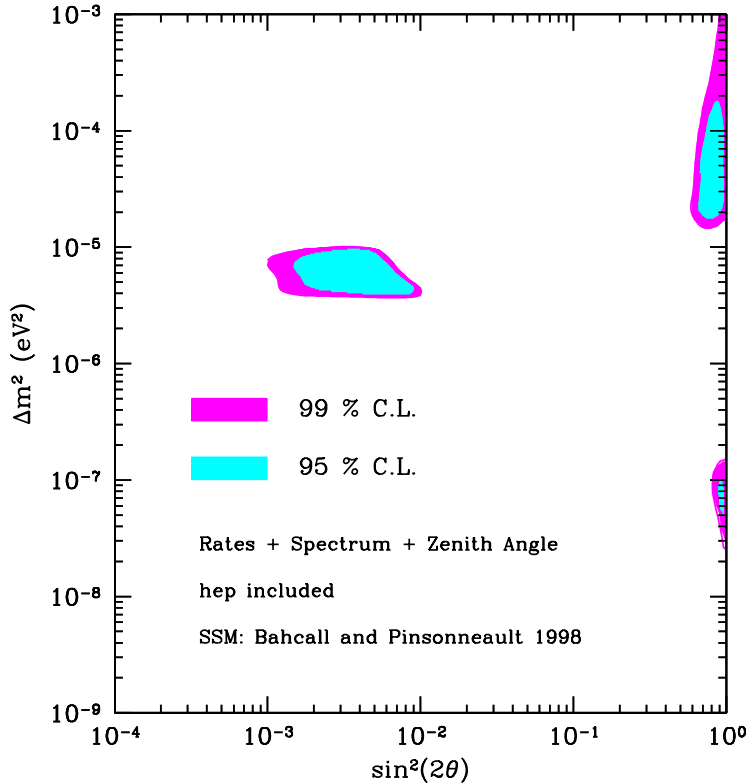,width=5in,angle=0}}
\caption{Global fits: MSW solutions. The figure
shows the regions in MSW parameter space that are
consistent with 
the total rates observed in  the four solar neutrino experiments (chlorine,
SuperKamiokande, GALLEX, and SAGE) and with the electron recoil 
energy spectrum and zenith angle distribution that are measured
by SuperKamiokande~\cite{superkamiokande504}. 
Contours are drawn at both $95$\% C.L. and $99$\% C.L.  .
\label{fig:global}}
\end{figure}

Figure~\ref{fig:global} shows the allowed ranges of MSW 
parameters for a global solution with arbitrary $hep$ flux. 
All three of the conventional MSW solutions~\cite{bks98}, 
small mixing angle(SMA), large mixing angle(LMA), and  low mass(LOW) neutrino
oscillations are allowed. In
the global MSW solution with the standard $hep$ flux, the LMA and LOW
solutions are marginally ruled out at $99$\% C.L. .

For vacuum oscillations, the value of $\alpha$ corresponding to the 
global $\chi^2_{\rm min}$ does not depend strongly on  $\Delta m^2$ and
$\sin^22\theta$ within the acceptable region.  
The improvement in the C.L. for acceptance
increases from $6$\% to $15$\% when an arbitrary $hep$ flux is considered.

The best-fit global MSW solution with an arbitrary $hep$ flux has
neutrino parameters given by 
$\Delta m^2   =   5.4\times 10^{-6} {\rm eV}^2$ 
and  $\sin^22\theta    =   5.0\times 10^{-3}$,
which are very close to the best-fit MSW parameters~\cite{bks98}  
with the standard 
(much smaller) $hep$ flux.
For vacuum oscillations, the best-fit global solution has
$\Delta m^2  =  7.8\times 10^{-11} {\rm eV}^2$ , and 
$ \sin^22\theta  = 0.71$,
again similar to the neutrino parameters for the best-fit vacuum
solution~\cite{bks98} with the standard $hep$ flux.

\section{Discussion}
\label{s:discussion}

We have calculated global fits to all the available solar neutrino
data allowing for an arbitrarily large $hep$ flux.  We find good
fits to all the  data, 
including the electron recoil energy
spectrum (see Fig.~\ref{fig:fitstospectrum})
reported by SuperKamiokande~\cite{superkamiokande504}.  The best
fits are obtained  
for $hep$ fluxes that are $\gsim 20$ times the flux predicted if the
best-available estimate(Eq.~[\ref{eq:hepvalue}])
 for the low-energy cross section factor for
the $hep$ reaction is used in the standard solar model calculations.
We have been unable to find an argument from first-principle physics
that rules out values of the cross section factor $S_0({\rm hep})$
that are as large as
required  by our best-fit solutions that are described in 
 Table~\ref{table:higherenergy}.

At first glance, these results seem discouraging. If one allows a large
$hep$ flux to account for the enhancement at higher energies of the
measured electron recoil energy spectrum~\cite{superkamiokande504},
then it would seem to be very difficult to infer anything fundamental
about neutrino physics 
from the measured recoil electron energy spectrum. After all, to a
good approximation the distortion of the spectrum can be represented
for small distortions by a single parameter~\cite{rosen} and 
we are suggesting that 
an additional (unknown) parameter be added to the
fit, namely, the magnitude of the $hep$ flux. 

Fortunately, 
the SuperKamiokande~\cite{superkamiokande504,superkamiokande300}, 
SNO~\cite{sno}, and
ICARUS~\cite{icarus} experiments can all test for the possible existence
of a large $hep$ flux by measuring  the energy spectrum beyond the
energy corresponding to the endpoint of the $^8$B neutrino spectrum.
Table~\ref{table:higherenergy} shows that  solutions with a large 
admixture of $hep$ neutrinos are expected to 
produce appreciable numbers of events more energetic than  $14$ MeV
in the 
$504$ days of observations studied so far in 
the SuperKamiokande detector.  
The region beyond the endpoint energy  of the $^8$B 
spectrum is an excellent region in which to search for rare events since the
background is expected to be very small between
$16$ to $20$ MeV .

The SNO detector should be even more sensitive than SuperKamiokande at the
highest electron energies because the neutrino absorption cross
section on deuterium rises more rapidly with energy than does the
electron scattering cross section and because higher energy neutrinos
absorbed by deuterium produce higher energy electrons, whereas for
$\nu-e$ scattering the energy is divided almost equally between
recoiling electrons and final state neutrinos~\cite{crosssections}. 
Quantitatively, we estimate that
SNO would  have, depending on which neutrino oscillation parameters are
chosen, two to three times the rate of production of electrons
in the $14-16$ MeV bin if the energy resolution were the same in the
two detectors.  Moreover, the energy discrimination for SNO 
may actually be
better than for SuperKamiokande, further helping in determining the
possible contribution of $hep$ neutrinos.

The ratio, $r$,  of the number of detected 
events in the $14-16$ MeV bin to the number
of detected events in the $16-20$ MeV bin should be large if---as
predicted by the standard solar model--- the 
only important neutrino sources contributing to events 
in this energy region are $^8$B and $hep$. 
For the best global fits (large $\alpha$), we see from 
Table~\ref{table:higherenergy},  that $r$ satisfies for
SuperKamiokande operating characteristics

\begin{equation} 
r({\rm global})
~\equiv~{  {({\rm events:} ~14-16 \ {\rm MeV})} \over {({\rm events:} ~16-20
 \ {\rm MeV})}  } ~~\gsim~~  4 \ , 
\label{eq:r1518}
\end{equation}
and for the standard $S_0({\rm hep})$, 
\begin{equation} 
r(\alpha = 1)
~\equiv~{  {({\rm events:} ~14-16 \ {\rm MeV})} \over {({\rm events:} ~16-20
 \ {\rm MeV})}  } ~~\gsim~~  10 \ .
\label{eq:ralpha1}
\end{equation}
Equation~(\ref{eq:r1518}) is a  prediction, 
valid with or without neutrino oscillations, of the standard solar model
and can be tested  with the available
SuperKamiokande~\cite{superkamiokande504} data.
Basically, Eq.~(\ref{eq:r1518}) is a statement that there are no other
important sources of high energy solar neutrino neutrinos except $^8$B
and $hep$. Equation~(\ref{eq:ralpha1}) is valid if the current
best-estimate for $S_0({\rm hep})$ is correct.

For $504$ days of SuperKamiokande operation, 
the best global fits predict (see Table~\ref{table:higherenergy}) 
about $12 \pm 2$ events in the
$16-20$ MeV energy bin, whereas the standard standard fluxes with 
 $\alpha = 1$ predict  $\sim 6$ or $7$ high energy events .  Many more
events may  be required before SuperKamiokande can  distinguish
empirically between the small and large $\alpha$ descriptions of the
energy spectrum. 

Measurements at energies below the current $6.5$ MeV lower limit are
also  very important. Figure~\ref{fig:fitstospectrum} shows that the
best-fit vacuum solution has a small upturn in the spectrum at the
lowest available energies. The upturn is 
intrinsic to  this vacuum solution;  $hep$ neutrinos are unimportant
at the lower energies. 

Solar neutrino experiments may be able to
determine, after several years of operation,  
both the contamination (at higher energies) 
by $hep$ neutrinos of the energy spectrum 
and also a strong constraint (from measurements at lower energies) 
on the allowed range of distortion
parameters due to neutrino oscillations.  
We hope that the discussion in this paper will stimulate 
further experimental and theoretical considerations of the possible
effects of $hep$ neutrinos. 

We are grateful to the SuperKamiokande collaboration for making
available at Neutrino 98 their crucially important data on the
electron recoil energy spectrum. We are also indebted to many members
of the SuperKamiokande collaboration for informative and stimulating
conversations on the possible relevance  of $hep$ neutrinos for their
measurements. 
This  work is an extension of our 
earlier collaboration with A. Smirnov, to whom we are grateful for many
insightful remarks,  on BKS98.
We are indebted to K. Babu, G. Beier, S. Bilenky, J. Carlson, D. Eisenstein, 
A. Gruzinov, W. Haxton, C. Kolda, H. Robertson, 
R. Shiavilla, and R. Wiring for
valuable comments that improved an initial version of this paper.
JNB and PIK acknowledge support from NSF grant \#PHY95-13835. PIK also
acknowledges support from \#PHY96-05140.

\end{document}